\documentclass{IEEEtran}
\usepackage{cite}
\usepackage{amsmath,amssymb,amsfonts}
\usepackage{algorithmic}
\usepackage{graphicx}
\usepackage{textcomp}
\usepackage{epstopdf}
\usepackage{xcolor}
\usepackage{todonotes}
\usepackage{microtype}
\usepackage{xparse}

\usepackage{tikz}
\usetikzlibrary{decorations.pathreplacing,calc}
\newcommand{\tikzmark}[1]{\tikz[overlay,remember picture] \node (#1) {};}
\newcommand*{\AddNote}[4]{%
    \begin{tikzpicture}[overlay, remember picture]
        \draw [decoration={brace,amplitude=0.5em},decorate,black]
            ($(#3)!(#1.north)!($(#3)-(0,1)$)$) --  
            ($(#3)!(#2.south)!($(#3)-(0,1)$)$)
                node [align=center, text width=2.5cm, pos=0.5, anchor=west] {#4};
    \end{tikzpicture}
}

\usepackage{algorithm}
\usepackage{algorithmic}
\usepackage[acronym]{glossaries}
\DeclareMathOperator*{\argmin}{argmin}
\DeclareMathOperator*{\argmax}{argmax}

\newacronym{EBA}{EBA}{Extrapolation-Based Algorithm}

\newacronym{iIRS}{IRS}{intelligent reflecting surface}
\newacronym{iCSI}{CSI}{Channel State Information}
\newacronym{iSNR}{SNR}{signal-to-noise ratio}
\newacronym{immWave}{mmWave}{millimeter wave}
\newacronym{iUT}{UT}{user tracking}

\newacronym{UT}{UT}{user tracking}
\newacronym{IRS}{IRS}{Intelligent reflecting surface}
\newacronym{CSI}{CSI}{channel state information}
\newacronym{UAV}{UAV}{Unmanned Aerial Vehicle}
\newacronym{LoS}{LoS}{line-of-sight}
\newacronym{ES}{ES}{Efficient Search Algorithm}
\newacronym{SNR}{SNR}{signal-to-noise ratio}
\newacronym{BS}{BS}{base station}
\newacronym[\glslongpluralkey={angles of arrival}]{AoA}{AoA}{angle of arrival}
\newacronym[\glslongpluralkey={angles of departure}]{AoD}{AoD}{angle of departure}
\newacronym{mmWave}{mmWave}{millimeter wave}
\newacronym{RF}{RF}{Radio Frequency}
\newacronym{DFT}{DFT}{Discrete Fourier Transform}
\newacronym{MSE}{MSE}{mean square error}
\newacronym{PoV}{PoV}{Point of View}
\newacronym{UPA}{UPA}{uniform planar array}
\newacronym{D}{D}{data transmission}
\newacronym{CE}{CE}{channel estimation}
\newacronym{IDE}{IDE}{IRS direction estimation}
\newacronym{CS}{CS}{compressed sensing}
\newacronym{SISO}{SISO}{single-input single-output}
\newacronym{NR}{NR}{New Radio}
\newacronym{OFDM}{OFDM}{Orthogonal Frequency-Division Multiplexing}
\newacronym{EKF}{EKF}{Extended Kalman Filter}
\newacronym{B1}{B1}{Baseline 1}
\newacronym{B2}{B2}{baseline 2}
\newacronym{FS}{FS}{full codebook search}
\newacronym{GLRT}{GLRT}{generalized likelihood ratio test}

\usepackage{cite}
\def\BibTeX{{\rm B\kern-.05em{\sc i\kern-.025em b}\kern-.08em T\kern-.1667em\lower.7ex\hbox{E}\kern-.125emX}}
\begin{document}
\title{Codebook-Based User Tracking in IRS-Assisted mmWave Communication Networks}

\author{Moritz Garkisch, Vahid Jamali, and Robert Schober\vspace{-0.5cm}}

\maketitle
\thispagestyle{plain}
\pagestyle{plain}

\begin{abstract}
In this paper, we present a novel mobile \gls{iUT} scheme for codebook-based \gls{iIRS}-aided \gls{immWave} systems. The proposed \gls{iUT} scheme exploits the temporal correlation of the direction from the \gls{iIRS} to the mobile user for selecting \gls{iIRS} phase shifts that provide reflection towards the user. To this end, the user's direction is periodically estimated based on a \gls{GLRT} and the user's movement trajectory is extrapolated from several past direction estimates. The efficiency of the proposed \gls{iUT} scheme is evaluated in terms of the average effective rate, which accounts for both the required signaling overhead and the achieved \gls{iSNR}. Our results show that for medium-to-high \gls{iSNR}, the proposed codebook-based \gls{iUT} scheme achieves a higher effective rate than two reference approaches based on full codebook search and optimization of the individual \gls{iIRS} unit cells, respectively.
\end{abstract}
\glsresetall

\vspace{-3mm}
\section{Introduction}
In the past years, \gls{mmWave} systems have been thoroughly analyzed, as they exploit previously unused spectrum and enable high data rates in wireless communication systems \cite{samimi2016mmwave}. For \gls{mmWave} frequencies, high path loss leads to few scatterers and a channel that is sparse in the angular domain, i.e., the received signal arrives from only few separable directions \cite{samimi2016mmwave}. \glspl{IRS} have been introduced to improve the channel gain for scenarios where the dominant \gls{LoS} path is blocked. By configuring the phase shifts of its unit cells properly, the \gls{IRS} can create a configurable propagation path \cite{estrakhri2016wavefront}. Configuring the \gls{IRS} requires \gls{CSI}, which is difficult to obtain in practice due to the typically large number of unit cells and the lack of sensing capabilities at the \gls{IRS}. As an alternative, resource-intensive channel estimation may be circumvented by configuring the \gls{IRS} based on predefined codewords from a phase shift codebook \cite{najafi2020physicsbased}.

For systems without \gls{IRS}, the problem of direction estimation has been discussed extensively in the literature, where sensing at the \gls{BS} antenna array is usually employed to infer information about the \gls{AoD} (see, e.g., \cite{va2016beamtracking}). However, these \gls{UT} schemes can not be employed in passive \gls{IRS}-assisted systems. A multiple \gls{IRS}-assisted \gls{UT} scheme for linear movement has been proposed in \cite{huang2022roadside}. However, to the best of the authors' knowledge, codebook-based \gls{IRS}-assisted \gls{UT} schemes for general non-linear user movement have not been reported, so far.

In this paper, we propose a novel \gls{UT} scheme, for \gls{IRS}-assisted wireless systems, that selects codewords from a predefined codebook to strengthen the \gls{LoS} link from the \gls{IRS} to the user. For user direction estimation, a \gls{GLRT} framework is proposed, and to minimize the signaling overhead, several past direction estimates are used to predict the user's future directions. We analyze the resulting effective rate of the system, which accounts for the tradeoff between signaling overhead and achievable \gls{SNR}. Based on simulations, the proposed \gls{UT} scheme is compared to two baseline schemes employing a full codebook search and optimization of the phase shifts of the \gls{IRS} unit cells based on full \gls{CSI}, respectively. Our results reveal that the proposed \gls{UT} scheme outperforms the full codebook search and the full \gls{CSI} baselines at medium-to-high \glspl{SNR}.

\textit{Notations}: Lower case and upper case bold letters denote vectors and matrices, respectively. The transpose and conjugate transpose of matrix $\mathbf{A}$ are denoted by $\mathbf{A}^{\mathrm{T}}$ and $\mathbf{A}^{\mathrm{H}}$, respectively. $\mathbf{I}_N$ is the identity matrix of size $N$. The $i$-th element of vector $\mathbf{a}$ is denoted by $[\mathbf{a}]_{i}$, and the element in the $i$-th row and $j$-th column of matrix $\mathbf{A}$ is denoted by $[\mathbf{A}]_{i,j}$. A complex normal distributed vector with mean vector $\mathbf{x}$ and covariance matrix $\mathbf{A}$ is represented by $\mathcal{CN}(\mathbf{x},\mathbf{A})$. Furthermore, the complex conjugate, absolute, and expected values of a scalar $x$ are denoted by $x^*$, $|x |$, and $\mathcal{E}\{x\}$, respectively. The $l^n$-norm is denoted by $\left\| \cdot \right\|_n$. The cardinality of set $\mathcal{M}$ is denoted by $|\mathcal{M}|$. The sets $\mathbb{N}$ and $\mathbb{C}$ denote natural and complex numbers, respectively. Finally, the big-O notation is denoted by $\mathcal{O}(\cdot)$.

\vspace{-1mm}
\section{System Model}
\begin{figure}
	\centering
    \includegraphics[clip, trim=3cm 6.8cm 2cm 6.7cm,width=0.35\textwidth]{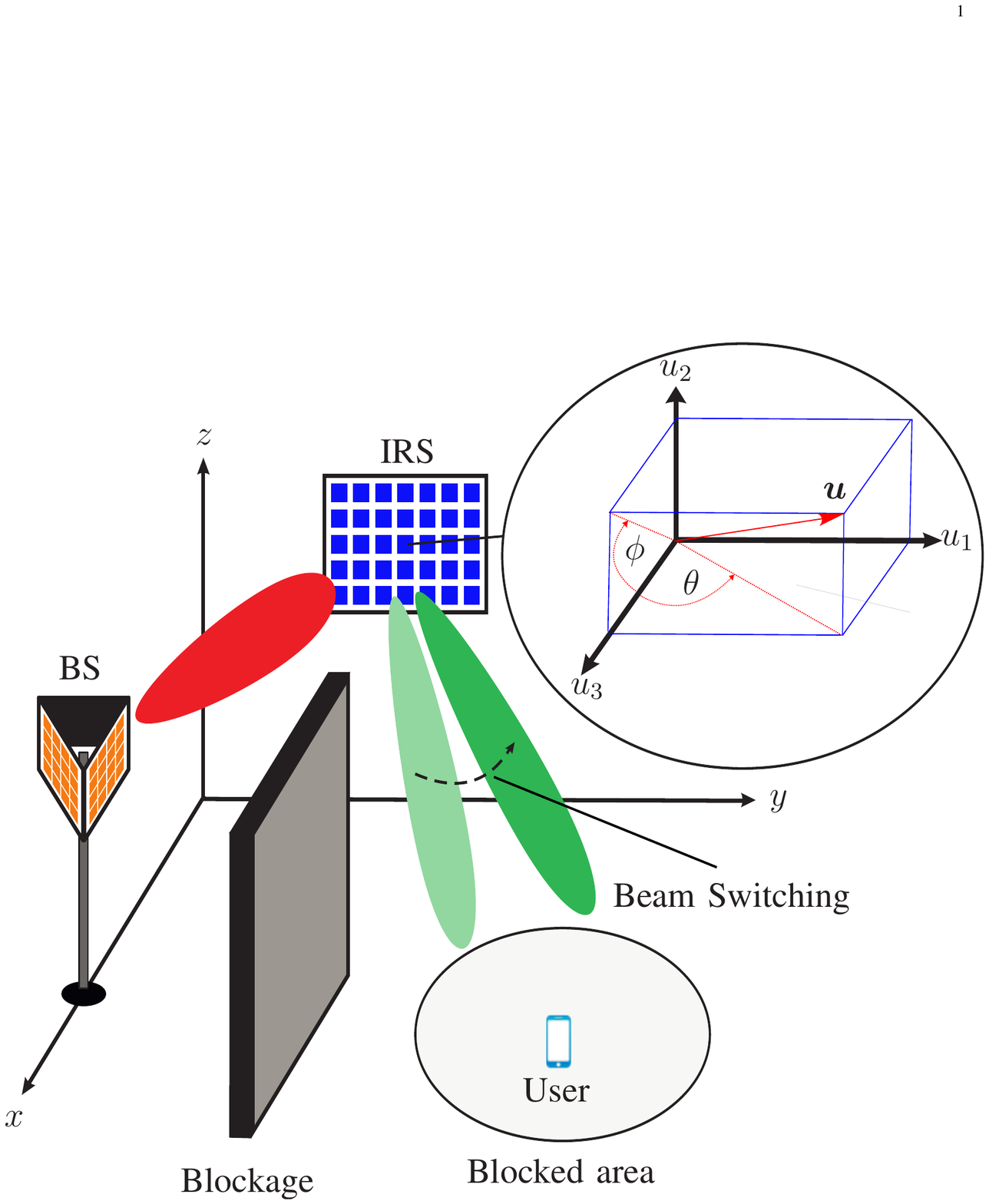}
    \vspace{-3mm}
    \caption{The considered system consists of a \gls{BS}, an \gls{IRS}, and a user that moves within an obstructed area. The direct link between \gls{BS} and user is blocked.} \label{fig:system_model}
    \vspace{-5mm}
\end{figure}

The considered system is illustrated in Fig. \ref{fig:system_model}. The coordinate system is defined by $[x,y,z]$. We consider a downlink system with one multi-antenna \gls{BS} equipped with $N_\mathrm{BS}$$=$$N_{\mathrm{BS},x}$$\times$$N_{\mathrm{BS},z}$ antennas arranged as an \gls{UPA} in the $x$-$z$ plane, and one single-antenna user. Furthermore, an \gls{IRS} consisting of $Q$$=$$Q_{y}$$\times$$Q_{z}$ unit cells lies in the $y$-$z$ plane, with unit cell area $A_\mathrm{UC}$$=$$d_{y}$$\times$$d_{z}$\footnote{For notational convenience, we assume that the \gls{BS} \gls{UPA} and the \gls{IRS} are located in the $x$-$z$ and $y$-$z$ planes, respectively. Nevertheless, adopting a more involved notation, the proposed \gls{UT} scheme can be adapted to general positions and orientations of \gls{BS} \gls{UPA} and \gls{IRS}.}. For an array lying in the $u_1$-$u_2$ plane, vector $\boldsymbol{u} = [u_1, u_2, u_3]^\mathrm{T}$ is parameterized by $\boldsymbol{\Psi}(\boldsymbol{u}) = [\theta, \phi]^\mathrm{T}$ with $\theta = \arctan(u_1/u_3)$ and $\phi = \arctan(u_2/u_3)$, for $u_3 > 0$. The direct link between the \gls{BS} and the user is assumed to be completely blocked. The positions of the centers of the antenna arrays at the \gls{BS} and \gls{IRS} are denoted as $\mathbf{p}_\mathrm{BS}$ and $\mathbf{p}_\mathrm{IRS}$, respectively. Throughout this paper, we assume that the positions of the \gls{BS} and \gls{IRS} are known, while the position of the user is generally unknown.\\

\vspace{-5mm}
\subsection{Codebook-Based Channel Model}
In this paper, we use a geometry-based transmission model that is defined by a limited number of propagation paths \cite{najafi2020physicsbased}. The received signal at the user is given by \cite{najafi2020physicsbased}
\begin{equation}\label{eq:mode_model_tx}
y(m,t) =  \mathbf{a}_{\mathrm{UE}} \mathbf{\Sigma}_{\mathrm{UE}} \mathbf{G}(m) \mathbf{\Sigma}_{\mathrm{BS}} \mathbf{D}_{\mathrm{BS}}^H \mathbf{f}(t) s(t) + n(t),
\end{equation}
where $\mathbf{a}_{\mathrm{UE}} = [a_1,...,a_{L_\mathrm{UE}}] \in \mathbb{C}^{L_\mathrm{UE}}$ contains unit norm scalars modeling the phases of the different reception paths at the user, $\mathbf{D}_{\mathrm{BS}} = [\mathbf{d}_1,...,\mathbf{d}_{L_\mathrm{BS}}] \in \mathbb{C}^{N_\mathrm{BS} \times L_\mathrm{BS}}$ contains the steering vectors for the \glspl{AoD} at the \gls{BS}, $\mathbf{\Sigma}_{\mathrm{UE}} \in \mathbb{C}^{L_\mathrm{UE} \times L_\mathrm{UE}}$ and $\mathbf{\Sigma}_{\mathrm{BS}} \in \mathbb{C}^{L_\mathrm{BS} \times L_\mathrm{BS}}$ are diagonal matrices containing the channel gains of the respective paths, and matrix $\mathbf{G}(m) \in \mathbb{C}^{L_\mathrm{UE} \times L_\mathrm{BS}}$ contains the \gls{IRS} response for all incoming to all outgoing directions for codeword $m$ from set $\mathcal{M}$, where $L_\mathrm{BS}$ and $L_\mathrm{UE}$ denote the numbers of paths in the \gls{BS}-to-\gls{IRS} and \gls{IRS}-to-user channels, respectively. By convention, we assume that $a_1$ and $\mathbf{d}_1$ correspond to their respective \gls{LoS} paths. We assume Rician fading, and the ratio of the power of the \gls{LoS} path to that of all other paths is denoted as $K = \frac{ [\mathbf{\Sigma}_i]_{1,1}^2 }{\sum_{k=2}^{L_i} [\mathbf{\Sigma}_i]_{k,k}^2 }$, $i\in\{ \mathrm{BS}, \mathrm{UE} \}$. Furthermore, $y(m,t)$, $\mathbf{f}(t) \in \mathbb{C}^{N_\mathrm{BS}}$, $s(t) \in \mathbb{C}$, and $n(t)\sim \mathcal{CN} (0,\sigma^2)$ represent the received signal for codeword $m$ at the user, the beamformer at the \gls{BS}, the transmit symbol, and additive white Gaussian noise with power $\sigma^2$, respectively. For simplicity, the transmit power $P_\mathrm{ TX}$ $=$ $\mathcal{ E}\{ |s(t)|^2\}$ is assumed to be identical for pilot and data symbols.
Furthermore, the transmit signal vector is set to beamform towards the \gls{IRS}, i.e., $\boldsymbol{f}(t) = \boldsymbol{f} = \boldsymbol{d}_1$, since the position of \gls{BS} and \gls{IRS} are known. Matrix entry $[\mathbf{G}(m)]_{i,j}$ is the \gls{IRS} response function $g_m(\mathbf{\Psi}_{\mathrm{BS}}, \mathbf{\Psi}_{\mathrm{UE}})$ of codeword $m$ for the $i$-th \gls{AoA} $\mathbf{\Psi}_\mathrm{BS}$ and the $j$-th \gls{AoD} $\mathbf{\Psi}_\mathrm{UE}$. The \gls{IRS} response function is given by \cite{estrakhri2016wavefront}
\begin{equation}
\small
\begin{split}
	& g_m(\mathrm{\mathbf{\Psi}_{\mathrm{BS}}}, \mathbf{\Psi}_{\mathrm{UE}}) = \\
	 & \bar{g} \sum_{q_{y}=0}^{Q_{y}-1} \sum_{q_{z}=0}^{Q_{z}-1} \mathrm{e}^{\mathrm{j}\frac{2\pi}{\lambda}\left( d_y A_{y} (\mathrm{ \mathbf{\Psi}_{\mathrm{BS}}}, \mathbf{\Psi}_{\mathrm{UE}})q_{y} + d_z  A_{z} (\mathrm{ \mathbf{\Psi}_{\mathrm{BS}}}, \mathbf{\Psi}_{\mathrm{UE}})q_{z} \right )} \mathrm{e}^{\mathrm{j} \omega_{q_{y},q_{z}}(m)  },
\end{split}
\end{equation}
where $\lambda$ is the wavelength, $\omega_{q_{y},q_{z}}(m)$ is the phase shift of the $(q_{y},q_{z})$-th unit cell, and $\bar{g}$ $=$ $\frac{4\pi A_\mathrm{UC}}{\lambda^2}$. Furthermore, we define $A_{y} (\mathrm{ \mathbf{\Psi}_{\mathrm{BS}}}, \mathbf{\Psi}_{\mathrm{UE}})$ $=$ $A_{y} (\mathrm{ \mathbf{\Psi}_{\mathrm{BS}}})$ $+$ $A_{y} (\mathbf{\Psi}_{\mathrm{UE}})$ and $A_{z} (\mathrm{ \mathbf{\Psi}_{\mathrm{BS}}}, \mathbf{\Psi}_{\mathrm{UE}})$ $=$ $A_{z} (\mathrm{ \mathbf{\Psi}_{\mathrm{BS}}})$ $+$ $A_{z} (\mathbf{\Psi}_{\mathrm{UE}})$ with $A_{y}(\mathbf{\Psi})$ $=$ $\sin(\alpha) \cos (\epsilon)$ and $A_{z}(\mathbf{\Psi})$ $=$ $\sin (\alpha) \sin (\epsilon)$, where $\alpha$ $=$ $\arctan\left( \sqrt{ \tan^2(\phi) + \tan^2(\theta)}\right)$ and  $\epsilon$ $=$ $\arctan \left( \tan(\theta) / \tan(\phi) \right) + \frac{\pi}{2} \left( 1-\mathrm{sign}(\tan(\phi) \right)$ \cite{jamali2020power}.

While the proposed \gls{UT} scheme is applicable for general \gls{IRS} phase-shift codebooks, for concreteness, we adopt the so-called quadratic codebook, which allows a flexible selection of the codebook size and beamwidth \cite{jamali2020power}. This codebook parameterizes the codewords $m$ by tuples $(m_{y}, m_{z})$, with $m_{y} \in \{ 0,...,M_{y}-1\}$, $m_{z} \in \{ 0,...,M_{z}-1\}$, and $M$=$M_{y}M_{z}$. In particular, the phase shift of unit cell $(q_{y},q_{z})$ for codeword $(m_{y},m_{z})$ is given as \cite{jamali2020power}
\begin{equation}
\begin{split}
	\omega_{q_{y},q_{z}}(m_{y},m_{z}) = & - \pi \left[ \frac{w\Delta\beta_{y,m_{y}}}{2Q_{y}}q_{y}^2 + \beta_{y,m_{y}} q_{y} \right]\\
	& - \pi \left[ \frac{w\Delta\beta_{z,m_{z}}}{2Q_{z}}q_{z}^2 + \beta_{z,m_{z}} q_{z} \right],
\end{split}
\end{equation}
where $\beta_{i,m_i+1} = \beta_{i,m_i} + \Delta \beta_i$, $\Delta \beta_i = \frac{2}{M_i}$, $\beta_{i,0} = -1$, for $i \in \{y,z\}$, and parameter $w \geq 0$ controls the beamwidth. In this paper, we employ $w=2$, which generates partially overlapping \gls{IRS} beams. Furthermore, we define the main lobe direction of the beam generated by codeword $m$, $\boldsymbol{\Psi}_\mathrm{IRS}(m) = (\theta_\mathrm{IRS}(m), \phi_\mathrm{IRS}(m))$, as the direction in which the highest reflection gain is achieved.

\vspace{-5mm}
\subsection{Transmission Block Structure}\label{sec:frame_model}
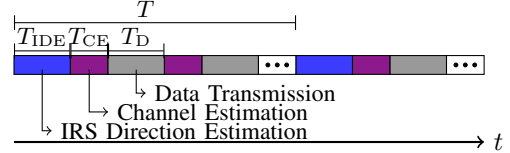
\begin{figure}
	\centering
	
	\begin{tikzpicture}[scale=0.25]
	\definecolor{myblue}{RGB}{60,60,255}
	
	\fill[myblue, draw=black] (1,0) rectangle (4,1);
	\draw[,->] (2.5,0.25) -- (2.5,-3) -- (3,-3) node[anchor=west] {\small IRS Direction Estimation};
	\fill[white!10!violet, draw=black] (4,0) rectangle (6,1);
	\draw[,->] (5,0.25) -- (5,-2.0) -- (6,-2.0) node[anchor=west] {\small Channel Estimation};
	\fill[gray!80!white, draw=black] (6,0) rectangle (9,1);
	\draw[,->] (7.5,0.25) -- (7.5,-1) -- (8,-1) node[anchor=west] {\small Data Transmission};
	\fill[white!10!violet, draw=black] (9,0) rectangle (11,1);
	\fill[gray!80!white, draw=black] (11,0) rectangle (14,1);
	
	\fill[white,draw=black] (14,0) rectangle (16,1);
	\fill[black, draw=black] (14.5,.5) circle (0.1cm);
	\fill[black, draw=black] (15,.5) circle (0.1cm);
	\fill[black, draw=black] (15.5,.5) circle (0.1cm);

    \draw [|-|](1,1.25) -- (4,1.25);
    \node at (2.5,2.0) {\small $T_\mathrm{IDE}$};
    \draw [|-|](4,1.25) -- (6,1.25);
    \node at (5,2.0) {\small $T_\mathrm{CE}$};
    \draw [|-|](6,1.25) -- (9,1.25);
    \node at (7.5,2.0) {\small $T_\mathrm{D}$};
    \draw [|-|](1,2.75) -- (16,2.75);
    \node at (8,3.5) {\small $T$};
    
    \fill[myblue, draw=black] (16,0) rectangle (19,1);
	\fill[white!10!violet, draw=black] (19,0) rectangle (21,1);
	\fill[gray!80!white, draw=black] (21,0) rectangle (24,1);
	
	\fill[white,draw=black] (24,0) rectangle (26,1);
	\fill[black, draw=black] (24.5,.5) circle (0.1cm);
	\fill[black, draw=black] (25,.5) circle (0.1cm);
	\fill[black, draw=black] (25.5,.5) circle (0.1cm);
    
    \draw[thick,->] (1,-3.6) -- (26,-3.6) node[anchor=west] {$t$};
    	\end{tikzpicture}
	\vspace{-5mm}
	\caption{Block diagram of the proposed frame structure.}
	\vspace{-3mm}
	\label{image:frame_structure}
	
\end{figure}

Before tracking the user, an initial connection from the \gls{BS} to the user via the \gls{IRS} has to be established, i.e. the codeword $\breve{m}(0)$ providing the largest reflection gain for the initial user position has to be found. In this paper, we focus on the tracking phase, since the establishment of the initial connection has been studied extensively in the literature, e.g., \cite{alexandropoulous2022nearfield}. The available time is divided into transmission blocks of equal length that start immediately after the initial connection, enumerated by $k\in\mathbb{N}$. Each transmission block starts with an \gls{IDE} frame, which is followed by $\eta \in \mathbb{N}$ alternating \gls{CE} and \gls{D} frames, see Fig. \ref{image:frame_structure}. The durations of the \gls{IDE}, \gls{CE}, and \gls{D} frames are denoted as $T_\mathrm{IDE}$, $T_\mathrm{CE}$, and $T_\mathrm{D}$, respectively. The data is transmitted during the \gls{D} frames and the \gls{CE} frames are used to estimate the end-to-end channel from the \gls{BS} to the user including the impact of the \gls{IRS}. The start time of transmission block $k$ is denoted by $t_k^{\mathrm{TB}}$, and the transmission block duration is $T = T_\mathrm{IDE} + \eta ( T_\mathrm{CE} + T_\mathrm{D} )$. During the \gls{CE} and \gls{D} frames, the \gls{IRS}-to-user direction is predicted based on previous direction estimates in the \gls{IDE} frames, which allows selecting an appropriate codeword from the codebook. For a mobile user, small-scale fading causes rapid changes to the channel parameters, whereas the direction changes only slowly. Therefore, the proposed \gls{UT} scheme operates on two time scales, where \gls{CE} has to be performed in short intervals as $T_\mathrm{CE}$+$T_\mathrm{D}$ can not exceed the channel coherence time, but \gls{IDE} can be performed infrequently \cite{alexandropoulous2022nearfield}. 

\vspace{-3mm}
\section{Codebook-Based Tracking Scheme}\label{sec:tracking}
In this section, we first propose a method to estimate the the user's direction. Subsequently, we extrapolate the movement trajectory for the entire following transmission block.

\vspace{-3mm}
\subsection{Direction Estimation}\label{sec:direction_measurement}
We assume that the \gls{IRS} is fully passive and cannot perform sensing. Therefore, we estimate the direction from the \gls{IRS} to the user, $\mathbf{ \Psi}_\mathrm{ UE}$, at the user and then feed it back to the \gls{BS} via a control channel. To this end, the \gls{IRS} cycles through several candidate codewords, while the \gls{BS} sends a pilot sequence of length $N_\mathrm{IDE}$ for each candidate codeword. The candidate codewords $m$ are chosen such that their main lobe directions, $\boldsymbol{\Psi}_\mathrm{IRS}(m_y, m_z)$,  are adjacent to 
the main lobe direction, $\boldsymbol{\Psi}_\mathrm{IRS}(\breve{m}_y, \breve{m}_z)$, of the currently employed codeword $\breve{m}$, and are defined by the set $(m_y,m_z) \in \mathcal{M}^{\mathrm{IDE}}$ where $\left\| (\breve{m}_y, \breve{m}_z) - (m_y, m_z) \right\|_\infty \leq \gamma$. Here, $\gamma \in \mathbb{N}$ is the maximum difference between the codeword indices. For simplicity, we assume that the pilot sequence is identical for all codewords and is denoted by $\mathbf{s}\in\mathbb{C}^{N_\mathrm{IDE}}$. We collect all measurements obtained for candidate codeword $m$ in vector $\mathbf{y}_m\in\mathbb{C}^{N_\mathrm{IDE}}$, such that $[\mathbf{y}_m]_j = y(m,t_m^{\mathrm{IDE}}+(j-1)T_\mathrm{S}), \;\;\forall m \in \mathcal{M}^\mathrm{IDE}$, where $T_\mathrm{S}$ denotes the symbol duration, $t_m^{\mathrm{IDE}}$ denotes the time when the \gls{IRS} is reconfigured with codeword $m$. For direction estimation, we simplify the system model in \eqref{eq:mode_model_tx}, since the \gls{LoS} path is dominant at \gls{mmWave} frequencies \cite{samimi2016mmwave} and the \gls{IRS} reflection codebook creates a narrow beam. Therefore, the received signal is approximated as follows\footnote{We note that the impact of scattering in the channel is included in our simulation results shown in Section \ref{sec:evaluation}.}
\begin{equation}\label{eq:channel_model_approx}
    y(m,t) \approx g_m(\boldsymbol\Psi_\mathrm{BS}, \boldsymbol\Psi_\mathrm{UE}) h s(t) + n(t),
\end{equation}
where $h = a_1 [\mathbf{\Sigma}_{\mathrm{UE}}]_{1,1} [\mathbf{\Sigma}_{\mathrm{BS}}]_{1,1} \mathbf{d}_{1}^H \mathbf{f}$ models the joint impact of the beamforming at the \gls{BS} and the \gls{LoS} channel. To estimate the direction of the user in the \gls{IDE} frame of transmission block $k$, we adopt the \gls{GLRT}-approach, which compares the likelihood of a set of hypotheses $\mathbf{\Psi}_\mathrm{UE} \in \mathcal{H}$, involving the unknown effective channel gain $h$ \cite{bethel2001amultihypothesis}, i.e.,
\begin{equation}\label{eq:original_direction_problem}
	\tilde{\boldsymbol{\Psi}}_\mathrm{UE}(k) = \argmax_{\boldsymbol{\Psi}_\mathrm{UE}\in \mathcal{H}} \max_{h} \prod_{m\in \mathcal{M}^{\mathrm{IDE}}} f(\boldsymbol{y}_m | \boldsymbol{\Psi}_\mathrm{UE} , h, \mathbf{s}),
\end{equation}
where $f(\mathbf{y}_m | \boldsymbol{\Psi}_\mathrm{UE} , h, \mathbf{s})$ is the conditional probability density function of $\mathbf{y}_m$. Here,  $f(\mathbf{y}_m | \boldsymbol{\Psi}_\mathrm{UE} , h, \mathbf{s})$ is complex Gaussian distributed with mean vector $g_m(\boldsymbol\Psi_\mathrm{BS}, \boldsymbol\Psi_\mathrm{UE}) h \mathbf{s}$ and covariance matrix $\sigma^2\mathbf{I}_{N_\mathrm{IDE}}$ and we assume that $h$ is constant because the \gls{IDE} frame is shorter than the channel coherence time. The inner maximization in \eqref{eq:original_direction_problem} over the effecting channel, $h$, that maximizes the likelihood of the observation, given a hypothesis $\mathbf{\Psi}_\mathrm{UE}$, can be obtained by setting the derivative of the objective function of \eqref{eq:original_direction_problem} with respect to $h$ to zero \cite{bethel2001amultihypothesis}. This yields
\vspace{-1mm}
\begin{equation}\label{eq:h_for_direction}
    \tilde{h}(\boldsymbol{\Psi}_\mathrm{UE}) = \frac{\sum_{m\in \mathcal{M}^{\mathrm{IDE}}} g_{m}^*(\boldsymbol\Psi_\mathrm{BS},\boldsymbol\Psi_\mathrm{UE}) \mathbf{s}^{\mathrm{H}} \mathbf{y}_m}{\sum_{m\in \mathcal{M}^{\mathrm{IDE}}} N_\mathrm{IDE} P_\mathrm{TX} | g_m(\boldsymbol\Psi_\mathrm{BS},\boldsymbol\Psi_\mathrm{UE})|^2 }.
\end{equation}
Now, according to the \gls{GLRT} principle \cite{bethel2001amultihypothesis}, we insert \eqref{eq:h_for_direction} into \eqref{eq:original_direction_problem}. Thus, \eqref{eq:original_direction_problem} can be equivalently reformulated as a non-linear least squares minimization problem:
\begin{equation}\label{eq:estimate_direction_final}
\small
	\tilde{\boldsymbol{\Psi}}_\mathrm{UE}(k) = \argmin_{\boldsymbol{\Psi}_\mathrm{UE}\in\mathcal{H}} \sum_{m\in\mathcal{M}^{\mathrm{IDE}}} \left \| \mathbf{y}_m-\tilde{h} (\boldsymbol{\Psi}_\mathrm{UE}) g_m(\boldsymbol{\Psi}_\mathrm{BS}, \boldsymbol{\Psi}_\mathrm{UE}) \mathbf{s}\right\|_2^2.
\end{equation}
Since we can not solve \eqref{eq:estimate_direction_final} in closed form, we evaluate \eqref{eq:estimate_direction_final} for the set of hypotheses $\mathcal{H} = \{ (\theta_1, \phi_1), ..., (\theta_H, \phi_H) \}, H\in\mathbb{N}$. For realistic human movement, the user's position is close to its last known position if the transmission block length is short, such that the set of hypotheses $\mathcal{H}$ is limited by the directions covered by the candidate codewords in $\mathcal{M}^{\mathrm{IDE}}$, where codeword $m'$ covers all directions $\mathbf{\Psi}_\mathrm{IRS}$ with $m' = \argmin_{m \in \mathcal{M}} \left \| \mathbf{ \Psi}_\mathrm{ IRS}(m) - \mathbf{ \Psi}_\mathrm{ IRS} \right\|_2$. Therefore, set $\mathcal{H}$ is defined by $\theta_n = \theta_\mathrm{IRS}(\breve{m}) + \left( n \frac{2\gamma+1}{H-1} -0.5-\gamma \right)\frac{180^\circ}{M_y}$ and $\phi_n = \phi_\mathrm{IRS}(\breve{m}) + \left( n \frac{2\gamma+1}{H-1} -0.5-\gamma \right)\frac{180^\circ}{M_z}, \; n\in\{ 0,...,H-1 \}$.

\vspace{-3mm}
\subsection{Extrapolation of the Trajectory}\label{sec:obtain_trajectory}
To reconfigure the employed \gls{IRS} codeword after the \gls{IDE} frame during the transmission block, we exploit the strong correlation between the past direction estimates and the future movement of the user. Thus, we extrapolate the user's trajectory from past direction estimates at the \gls{BS}. To this end, we assume that the user's movement direction is a smooth function of time over long periods, i.e., at least for several seconds. This property has been exploited in \cite{zhang2020codebook} and \cite{stratidakis2020low} to predict the future position of a user in codebook-based \gls{UT} schemes for wireless systems without \gls{IRS}, by linear extension of three past position measurements. We extend this concept to \gls{IRS}-assisted systems by fitting an arbitrary number of past estimates to an $n$-th order polynomial:
\vspace{-1mm}
\begin{equation}\label{eq:general_polynomial}
    \widehat{\boldsymbol\Psi}_\mathrm{UE}(t) = \sum_{i=0}^n \boldsymbol{c}_i t^i,
\end{equation}
\vspace{-0mm}
where $\widehat{\boldsymbol\Psi}_\mathrm{UE}(t) = [\hat\theta_\mathrm{UE}(t) , \hat\phi_\mathrm{UE}(t)]^{\mathrm{T}}$ is the predicted direction at time $t$, and $\boldsymbol{c}_i = [c_{i,\theta} , c_{i,\phi}]^\mathrm{T} \;\forall i$, are the weights of the polynomial. We adopt the direction estimates obtained in the last $S$ \gls{IDE} frames, c.f. \eqref{eq:estimate_direction_final}, for curve fitting\footnote{During the first few transmission blocks, the number of past direction estimates and the order of the polynomial have to be reduced.} and minimize the \gls{MSE} between the trajectory and the direction estimates:
\vspace{-1mm}
\begin{equation}\label{eq:min_mse}
	\min_{\boldsymbol{c}_{0},...,\boldsymbol{c}_{n}} \;\;\;   \frac{1}{S} \sum_{k=k'-S+1}^{k'} \left\| \hat{\boldsymbol{\Psi}}_\mathrm{UE}(t_k^{\mathrm{TB}}) - \tilde{\boldsymbol{\Psi}}_\mathrm{UE}(k) \right\|^2_2,
\end{equation}
\vspace{-1mm}
where the current \gls{IDE} frame is denoted by $k'$. Equation \eqref{eq:min_mse} can be solved analytically by setting the derivatives of the polynomial weights, $\mathbf{c}_i$, to zero. We omit the resulting expression due to space constraints. The polynomial weights are computed in every \gls{IDE} frame after direction estimation before the start of the following \gls{CE} frame. Thus, at the beginning of each \gls{CE} frame, i.e., at times $t'= t_{k'}^{\mathrm{TB}} + T_\mathrm{IDE} + i_\mathrm{CE} (T_\mathrm{CE}+T_\mathrm{D})$, $i_\mathrm{CE}=0,...,\eta-1$, the \gls{BS} uses the extrapolated trajectory to determine the codeword, whose main lobe direction is closest to $\hat{\boldsymbol{\Psi}}_\mathrm{UE}(t)$, as follows
\begin{equation}\label{eq:select_best_m}
    \breve{m}(t') = \argmin_{m \in \mathcal{M}} \left \| \hat{\boldsymbol{\Psi}}_\mathrm{UE}(t') - \boldsymbol{\Psi}_\mathrm{IRS}(m) \right \|_2^2.
\end{equation}

The proposed \gls{UT} scheme is summarized in Algorithm \ref{al:prediction_algorithm}.
\begin{algorithm}[]
\caption{Proposed User-tracking Scheme}
\begin{algorithmic}[1]
\footnotesize
    \REQUIRE Initial codeword $\breve{m}(0)$.
	\REPEAT
		\STATE Estimate direction according to \eqref{eq:h_for_direction} and \eqref{eq:estimate_direction_final}. \tikzmark{top} \tikzmark{right}
		\STATE Fit trajectory according to \eqref{eq:min_mse}.\tikzmark{bottom}
		\STATE \textbf{repeat} $\eta$ \textbf{times}
            \STATE \;\;\;\; Select appropriate codeword according to \eqref{eq:select_best_m}.
	        \STATE \;\;\;\; End-to-end channel estimation (\gls{CE} frame).
	        \STATE \;\;\;\; Data transmission (\gls{D} frame).
	    \STATE \textbf{end} \textbf{repeat}
\UNTIL{User exits obstructed area.}
\vspace{-5mm}
\end{algorithmic}\label{al:prediction_algorithm}
\AddNote{top}{bottom}{right}{\gls{IDE} frame}
\end{algorithm}
\setlength{\textfloatsep}{3pt}
 
\vspace{-3mm}
\subsection{Remarks on the Overhead}\label{sec:overhead}
The main advantage of the proposed \gls{UT} scheme is its low overhead, which is quantified in the following. As explained in Section \ref{sec:frame_model}, in each transmission block, one \gls{IDE} frame and $\eta$ \gls{CE} frames are needed in addition to the $\eta$ \gls{D} frames. The percentage of time needed for those frames, i.e., the resulting overhead, is $\Gamma = \frac{T_\mathrm{IDE} + \eta T_\mathrm{CE} }{ T }$, where the durations for testing all candidate codewords and \gls{CE} are $T_\mathrm{IDE} = |\mathcal{M}^{\mathrm{IDE}}| N_\mathrm{IDE} T_\mathrm{S}$ and $T_\mathrm{CE}=N_\mathrm{CE}T_\mathrm{S}$, respectively, where $N_\mathrm{CE}$ is the number of pilot symbols for \gls{CE}. For recovery, neglecting noise, $N_\mathrm{CE}$ needs to be at least as large as the number of channel coefficients to be estimated. For the proposed \gls{UT} scheme, $h$ is a scalar and thus, the number of pilot symbols for \gls{CE} does not scale with the \gls{IRS} size, e.g., $N_\mathrm{CE} \sim \mathcal{O}(1)$. In contrast to our approach, most existing schemes in the literature require full \gls{CSI}, comprising the individual channels of each \gls{IRS} unit cell, and design the \gls{IRS} phase shifts for one subsequent \gls{D} frame. This yields a signaling overhead of $\Gamma = \frac{T_\mathrm{CE}^\mathrm{B}}{ T_\mathrm{CE}^\mathrm{B} + T_\mathrm{D}^\mathrm{B} }$, with $T_\mathrm{CE}^\mathrm{B} = N_\mathrm{CE}^\mathrm{B} T_\mathrm{S}$, where $N_\mathrm{CE}^\mathrm{B}$ and $T_\mathrm{D}^\mathrm{B}$ denote the corresponding pilot sequence length and the duration of the \gls{D} frame, respectively. To estimate the channel efficiently, \gls{CS} schemes have been proposed that exploit the sparsity of the \gls{mmWave} channel. In this case, the required number of pilot symbols scales with $N_\mathrm{CE}^{\mathrm{B}} \sim \mathcal{O}(L_\mathrm{BS}L_\mathrm{UE}\ln(Q))$ \cite{wang2020compressed}.

\section{Performance Evaluation}\label{sec:evaluation}
\begin{table}[]
\centering
\vspace{-3mm}
\caption{Simulation Settings. } \label{tab:general_settings}
\vspace{-3mm}
\scalebox{0.86}{
\begin{tabular}{|cc||cc||cc|}
\hline
$\mathbf{p}_\mathrm{BS}$    & $[0,0,10]$ m   & $Q_{y}, Q_{z}$                   & 100, 100    & $T$ & 0.15 s                       \\ \hline
$\mathbf{p}_\mathrm{IRS}$   & $[-40,40,5]$ m & $d_{y}, d_{z}$                   & $\lambda/2$ & $T_\mathrm{CE}+T_\mathrm{D}$                  &   1.29 ms                   \\ \hline
$r, r_\mathrm{C}$ & 15 m, 7.5 m  & $N_\mathrm{BS}$   & 12x4  & $T_\mathrm{S}$                & 4.16 $\mu$s \\ \hline
$v$                          & 5 km/h       & $L_\mathrm{BS}$, $L_\mathrm{UE}$ & 4, 4        & $K$                       &   3                   \\ \hline
$N_\mathrm{IDE}$           &    3        &    $S$                          &  3         &      $\sigma^2$                &       - 120 dBm            \\ \hline
$|\mathcal{M}^{\mathrm{IDE}}|$        &  9        &   $n, \gamma$                &    1,1  &   $f_\mathrm{c}$         &    28 GHz                     \\ \hline
\end{tabular}
}
\end{table}

For the following evaluation, we use a movement model similar to the one in \cite{dehkordi2021adaptive}, where the user moves with constant speed $v$ inside a circle of radius $r$. The movement involves three stages. First, the user enters the circle from a random angle and moves towards the center. When reaching a distance of $r_\mathrm{C}$ to the center, the user follows a circular trajectory around the center in counter-clockwise direction. Finally, at a random angle, the user moves straight away from the center and leaves the circle. The simulation parameters are collected in Table \ref{tab:general_settings}. We consider two baseline schemes\footnote{We cannot compare with the \gls{UT} scheme in \cite{huang2022roadside}, since it assumes linear movement and employs an algorithm that is not computationally feasible for large \gls{IRS}.}. The first baseline employs the frame structure presented in Section \ref{sec:frame_model} and performs a \gls{FS} to select the codeword yielding the largest received power for the following transmission block. The second baseline is identical to Baseline 3 from \cite{alexandropoulous2022nearfield}, i.e., it assumes perfect \gls{CSI} and optimizes the phase-shifts of the individual unit cells. We refer to this scheme as "full optimization baseline". Similar to the proposed \gls{UT} scheme, for both baselines, the \gls{BS} focuses on the \gls{LoS} towards the \gls{IRS}. In the following, we analyze the effective rate, defined as $R(t)$=$(1-\Gamma) \log_2(1+\mathrm{SNR}(t))$, averaged over the entire movement process, which shows the tradeoff between achieved \gls{SNR}, defined as $\mathrm{SNR}(t)$=$\frac{|y(m^*(t),t)|^2}{\sigma^2}$, and the required signaling overhead $\Gamma$. Determining the signaling overhead is non-trivial since it depends on the specific parameters of the respective system setup. Hence, we employ a first order approximation of the signaling overhead to render the effective rate of the different approaches comparable. To this end, we approximate the pilot symbol lengths for channel estimation by their scaling orders, i.e., $N_\mathrm{CE}$=$1$ for the proposed \gls{UT} scheme and \gls{FS} and $N_\mathrm{CE}^{\mathrm{B}}$=$L_\mathrm{BS} L_\mathrm{UE} \mathrm{ln}(Q) $$\approx$$147$ for the full optimization baseline, c.f. Table \ref{tab:general_settings}. 

\begin{figure}
	\centering
    \includegraphics[clip, trim=0cm 0cm 0cm 0cm,width=0.48\textwidth]{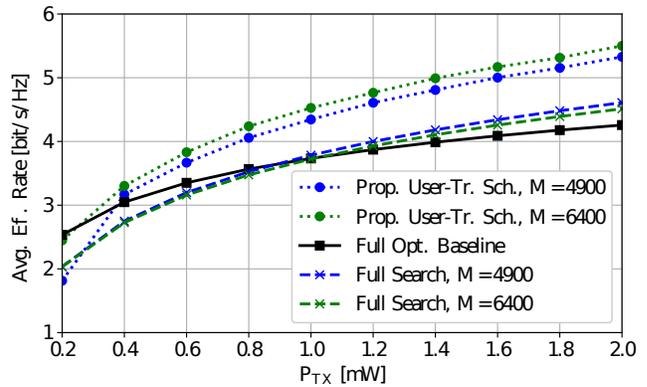}
    \vspace{-5mm}
	\caption{Effective rate versus transmit power.}\label{image:rate_power}
\end{figure}

The effective rate as a function of the \gls{BS} transmit power $P_\mathrm{TX}$ is shown in Fig. \ref{image:rate_power}. The results show that our proposed \gls{UT} scheme outperforms \gls{FS} for medium-to-high transmit powers. In this range, the \gls{IRS} reflection gain achieved by both schemes is similar, since both schemes employ the same codebook and erroneous codeword selection rarely occurs, but the overhead of \gls{FS} is larger leading to a lower effective rate. In the low power regime, inaccurate direction estimation may cause the proposed \gls{UT} scheme to select codewords that provide a sub-optimal reflection gain. This has a less severe impact on \gls{FS} as it cycles through the entire codebook in each \gls{IDE} frame. Furthermore, Fig. \ref{image:rate_power} shows that for \gls{FS}, the higher reflection gain enabled by a larger codebook ($M$=$6400$) does not compensate for the resulting higher overhead, such that a smaller codebook ($M$=$4900$) achieves a higher effective rate. The full optimization of the \gls{IRS} phase shifts constitutes an upper bound for the achievable reflection gain of the \gls{IRS}, but the required overhead is high. Thus, the significantly lower signaling overhead of the proposed \gls{UT} scheme leads to a higher effective rate in medium-to-high transmit power regime, since the signaling overhead affects the rate linearly, while the \gls{SNR} affects the rate only logarithmically.

\vspace{-3mm}
\section{Conclusion}
In this paper, a novel \gls{UT} scheme for codebook-based \gls{IRS}-assisted systems was introduced. The proposed \gls{UT} scheme exploits the high temporal correlation of a moving user's direction, by regularly estimating the user's direction based on a \gls{GLRT} approach and subsequently extrapolating the user's movement trajectory. Our simulation results have revealed that for sufficiently high \gls{SNR} the proposed scheme achieves a significantly higher effective rate than two baseline schemes from the literature.

\bibliographystyle{IEEEtran}
\bibliography{thesis.bib}

\end{document}